\documentclass[prl,showpacs,hyperref,superscriptaddress,twocolumn]{revtex4}
\usepackage{amsfonts}
\usepackage{amsmath}
\usepackage{amssymb}
\usepackage{graphicx}
\usepackage{subfigure}
\providecommand{\U}[1]{\protect\rule{.1in}{.1in}}
\providecommand{\U}[1]{\protect\rule{.1in}{.1in}}

\begin{document}

\title{Classical bifurcation in a quadrupolar NMR system}

\author{A. G. Araujo-Ferreira} \email{avatar@ifsc.usp.br}
\affiliation{
Instituto de F\'{\i}sica de S\~{a}o Carlos, Universidade de S\~{a}o Paulo, Caixa Postal 369, 13560-970 S\~{a}o Carlos, S\~{a}o Paulo, Brazil}
\author{R. Auccaise}
\affiliation{
Empresa Brasileira de Pesquisa Agropecu\'{a}ria, Rua Jardim Bot\^{a}nico 1024, 22460-000 Rio de Janeiro, Rio de Janeiro, Brazil}
\author{R. S. Sarthour}
\affiliation{
Centro Brasileiro de Pesquisas F\'{\i}sicas, Rua Dr. Xavier Sigaud 150, 22290-180 Rio de Janeiro, Rio de Janeiro, Brazil}
\author{I. S. Oliveira}
\affiliation{
Centro Brasileiro de Pesquisas F\'{\i}sicas, Rua Dr. Xavier Sigaud 150, 22290-180 Rio de Janeiro, Rio de Janeiro, Brazil}
\author{T. J. Bonagamba}
\affiliation{
Instituto de F\'{\i}sica de S\~{a}o Carlos, Universidade de S\~{a}o Paulo, Caixa Postal 369, 13560-970 S\~{a}o Carlos, S\~{a}o Paulo, Brazil}
\author{I. Roditi}
\affiliation{
Centro Brasileiro de Pesquisas F\'{\i}sicas, Rua Dr. Xavier Sigaud 150, 22290-180 Rio de Janeiro, Rio de Janeiro, Brazil}

\begin{abstract}
The Josephson Junction model is applied to the experimental implementation of classical bifurcation in a quadrupolar Nuclear Magnetic Resonance system. There are two regimes, one linear and one  nonlinear which are implemented  by the radio-frequency term and the quadrupolar term of the Hamiltonian of a spin system respectively. Those terms provide an explanation of the symmetry breaking due to bifurcation. Bifurcation depends on the coexistence of both regimes at the same time in different proportions.  The experiment is performed on a lyotropic liquid crystal sample of an ordered ensemble of $^{133}$Cs nuclei with spin $I=7/2$ at room temperature. Our experimental results confirm that bifurcation happens independently of the spin value and of the physical system. With this experimental spin scenario, we confirm that a quadrupolar nuclei system could be described analogously to a symmetric two--mode Bose--Einstein condensate.
\end{abstract}

\maketitle

The Josephson Junction (JJ) model remains one of the key concepts for theoretical advances in the physics of superconductivity and superfluidity. Within the ultracold atom scenario, the description of two--mode Bose--Einstein condensates (BEC) by means of the JJ model has afforded new insights into nonlinear tunneling \cite{albiez2005,gati2007}, owing to nonlinearity as a source, revealing many kinds of phenomena, from entanglement to classical bifurcation \cite{raghavan1999,sorensen2001May,kellman2002,hines2005,gsantos, angelajpa, angelaahp, zibold2010}. Lately, classical bifurcation effects have attracted the attention of researchers, as they can indicate, for the associated quantum systems, a signature of quantum phase transitions \cite{shchesnovich2009,diaz2010Feb,diaz2010Jun,rubeni2012} and, more recently, have been used in the study of an unstable quantum pendulum \cite{gerving2012}.  In this context and due to the huge advances in experimental control, the investigation of systems such as two BEC traps \cite{albiez2005,kellman2002,zibold2010} or two vibrational degrees of freedom in polyatomic molecules \cite{lehmann1983,kellman1985,li1990} has become very active. The Hamiltonians modeling such systems are well described by raising and lowering operators and they may be rewritten in terms of the SU(2) operators and their commutation relations via the Schwinger pseudospin representation. In this representation, these Hamiltonians display nonlinear terms that commonly appear in the $z$--component, as for instance in the case of a symmetric trap of a two--mode BEC. One may write the simplest Hamiltonian model as $\mathcal{H} = \chi \textbf{J}_{z}^{2}-\Omega \textbf{J}_{x}$, where $\chi$ and $\Omega$ represent respectively  the nonlinear coupling due to atom--atom interaction and the linear coupling due to an external perturbation. By this and similar models, some efforts have been made to understand the evolution of the wave function $\left|\Psi\left(t\right)\right\rangle$  in the spin--coherent representation, and to analyze entanglement \cite{sorensen2001May}, chaos in kicked spin systems \cite{nakamura1986} and bifurcation \cite{smerzi1997}.

The Hamiltonian $\mathcal{H}$, in the Nuclear Magnetic Resonance (NMR) scenario, has a direct physical interpretation and in the rotating frame  corresponds to the Hamiltonian of a quadrupolar nuclear system acted on by a radio-frequency pulse along the negative $x$--direction \cite{slichter1992,oliveira2007}. In this letter, we explore this equivalence of interpretation and study the feasibility of observing a signature of bifurcation in quadrupolar spin systems, while also assessing the use of the JJ--model in nuclear systems.

Nuclear spin systems with $I>1/2$ are described by the Zeeman term, the quadrupolar term, the radio--frequency term and  weak interactions with, among other things, nuclei, electrons and field fluctuations, which we refer to as an {\it environment term} denoted by $\mathcal{H}_{env}$ \cite{slichter1992}.

The Zeeman term  is the interaction between the spin nuclear magnetic moment $-\hbar\gamma \mathbf{I}=-\hbar\gamma \left(\mathbf{I}_{x},\mathbf{I}_{y},\mathbf{I}_{z}\right) $ and a strong constant magnetic field $\mathbf{B}_{0}=\left(0,0,B_{0}\right)$ aligned in the $z$--direction. The quadrupolar term is due to the interaction of the nuclear quadrupole moment ($Q$) with an electric field gradient ($V_{\alpha\beta}$, with $\alpha,\beta=x,y,z$), such that in an oriented system with axial symmetry, the following inequality is satisfied $
\left| V_{zz}\right| \gg \left| V_{xx}\right| \approx \left| V_{yy}\right|$, which allows us to express the term as $\frac{eQV_{zz}}{4I\left( 2I-1\right) }\left(  3\mathbf{I}_{z}^{2}-\mathbf{I}^{2}  \right) $. The radio--frequency (\textit{RF}) term represents the interaction between the spin nuclear magnetic moment and an external time--dependent magnetic field, which is perpendicular to the strong constant magnetic field $\mathbf{B}_{0}$; this term is written as $+\hbar \gamma B_{1}\left( \mathbf{I}_{x}\cos \left(\omega _{RF}t+\phi \right) +\mathbf{I}_{y}\sin \left( \omega _{RF}t+\phi\right) \right) $, where the phase $\phi $ defines its direction on the $xy$--plane.  In a rotating frame, the NMR Hamiltonian is given by

\begin{eqnarray}
\mathcal{H}_{NMR} &=&-\hbar \left( \omega _{L}-\omega _{RF}\right) \mathbf{I}%
_{z}+\hbar \frac{\omega _{Q}}{6}\left( 3\mathbf{I}_{z}^{2}-\mathbf{I}%
^{2}\right)  \nonumber \\
&&+\hbar \omega _{1}\left( \mathbf{I}_{x}\cos \phi +\mathbf{I}_{y}\sin
\phi \right) + \mathcal{H}^{\prime}_{env} \text{,}  \label{hamiltonianoRMNQ}
\end{eqnarray}
where $\omega _{Q}=\frac{3eQV_{zz}}{2I\left( 2I-1\right) \hbar }$ represents the quadrupolar coupling, $\omega _{1}=\gamma B_{1}$ is the \textit{RF} strength, and $\omega _{L}=\gamma B_{0}$ the Larmor frequency of a nuclear species. The Larmor frequency and quadrupolar coupling satisfy the inequality $ \left| \omega _{L}\right| \gg \left| \omega _{Q}\right|$.

\begin{figure}[tbp]
\includegraphics[width=0.48\textwidth]{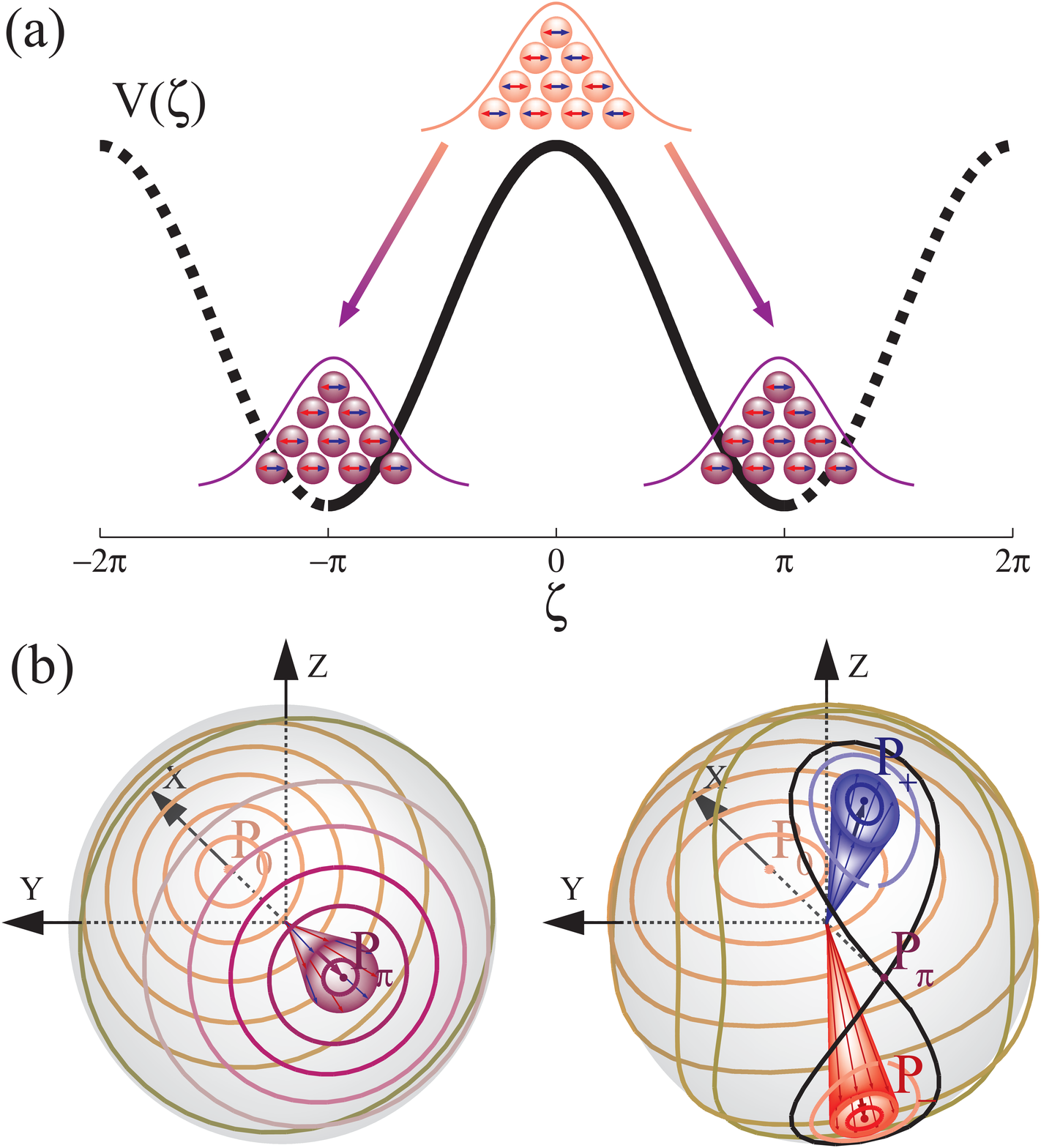}
\caption{(Color online) Sketch of the spin scenario of the bifurcation process discussed in this letter. (a) Initially the spin state precesses around fixed point P$_{0}$, such that after the dynamics the spin system goes to one of the points P$_{\pm}$ depending on the value of $\Lambda$.  (b) Trajectories drawn on the spherical phase space before and after the bifurcation process that depends on the value of $\Lambda$. The typical supercritical pitchfork bifurcation scenario occurs; i.e., a stable fixed point bifurcates into two new stable fixed points while the original becomes unstable.}
\label{fig:Fundamentals}
\end{figure}

To match the Hamiltonian $\mathcal{H}$ and the NMR Hamiltonian $\mathcal{H}_{NMR}$, let us choose  $\omega_{RF}=\omega _{L}$ and $\phi =\pi$, such that, without loss of generality, we may drop the constant term $-\frac{\hbar \omega _{Q}}{6}\mathbf{I}^{2}$ and the environment term $ \mathcal{H}^{\prime}_{env} $. Now the NMR Hamiltonian takes the form $\mathcal{H}_{NMR} = \frac{\hbar\omega _{Q}}{2} \mathbf{I}_{z}^{2}  -\hbar \omega _{1} \mathbf{I}_{x}  $.  Next, by substituting the dimensionless parameter $\Lambda= \frac{I  \omega _{Q}}{   \omega _{1}}$ the NMR Hamiltonian can be rewritten:

\begin{eqnarray}
\mathcal{H}_{NMR}^{\prime}&=&  \frac{\mathcal{H}_{NMR}}{\hbar \omega _{1}} =   \frac{\Lambda}{2  I}\mathbf{I}_{z}^{2}  - \mathbf{I}_{x} \text{.} \label{squeezingRMNQ} 
\end{eqnarray}
We then use this Hamiltonian to look into the classical bifurcation mechanism in nuclear spin systems. The corresponding semiclassical Hamiltonian is generated by mapping the quantum mechanical operators onto the complex numbers, following the usual procedure reported in \cite{smerzi1997,raghavan1999}. This amounts to letting $\mathbf{I}_{z}\rightarrow z$ and $\mathbf{I}_{x}  \rightarrow \sqrt{1-z^2}\cos \zeta$, giving

\begin{eqnarray}
\mathcal{H}^{\prime}&=&    \frac{\Lambda}{2} {z}^{2}  - \sqrt{1-z^2}\cos \zeta \text{,}  \label{HamiltonianoClassico} 
\end{eqnarray}
where, for our purposes, $z$ represents the temporal mean $z$--magnetization and $ \zeta $ a relative phase. In classical mechanics, Eq. (\ref{HamiltonianoClassico}) describes the motion of a particle in a phase potential defined by $V\left(\zeta\right)=-\sqrt{1-z^2}\cos \zeta $, where $V$ is shaped by $\cos \zeta$ and weighted by  $\sqrt{1-z^2}$ as sketched in Fig. \ref{fig:Fundamentals}(a). $\Lambda$  is a tunable parameter that determines the dynamics of a particle in a conserved energy configuration. From Hamilton's equations of motion $\dot{z}=-\partial \mathcal{H}^{\prime} / \partial \zeta $ and $\dot{\zeta}=\partial \mathcal{H}^{\prime} / \partial z $ we have,
\begin{eqnarray}
\dot{z}&=& - \sqrt{1-z^2}\sin \zeta \text{,} \label{EquacaoMovim1} \\
\dot{\zeta}&=& \Lambda z+\frac{z}{\sqrt{1-z^2}} \cos \zeta \text{,} \label{EquacaoMovim2} 
\end{eqnarray}
such that $z$ and $\zeta$ are canonically conjugate variables. The fixed points of the Hamiltonian $\mathcal{H}^{\prime}$, denoted by $P=\left(z_{0},\zeta_{0}\right)$,  are determined by the condition $\dot{z} = \dot{\zeta} = 0$. These are $\zeta_{0} =\pm n \pi$ with $n=0,1,2,\ldots$ and 
$z_{0}=\left\{0 , \pm \sqrt{1-1/\Lambda^2}\right\}$. Note that there are many interesting sets of fixed points to analyze. The trivial one, $P_{0}=\left(0,0\right)$, which corresponds to a stable fixed point,  is located on the positive $X$--axis of the coordinate frame (see Fig \ref{fig:Fundamentals}(b)).  Physically, in the NMR scenario, it corresponds to an extreme situation of null quadrupolar coupling, or to a linear regime ($\Lambda < 1$). Still in the linear regime, $P_{\pi}=\left(0,\pi\right)$ matches another stable fixed point, located on the negative X--axis of the coordinate frame  (see Fig \ref{fig:Fundamentals}(b)). The non-trivial fixed points are  $P_{\pm}=\left(\pm \sqrt{1-1/\Lambda^2},\pi\right)$; here, for $\Lambda > 1$, it can be seen that the stable fixed point P$_{\pi}$ undergoes a supercritical Pitchfork bifurcation, becoming unstable, and it splits up into two others, the $P_{\pm}$ stable fixed points. These are divided by a separatrix in Fig. \ref{fig:Fundamentals}(b). In the NMR interpretation, this picture corresponds to the situation of a quadrupolar coupling stronger than the $RF$ pulse intensity, or to a nonlinear regime. Our efforts were focused on finding out how this theoretical analysis could be reached by using nuclear spin systems.

The spin scenario and the experimental implementations were performed in a lyotropic liquid crystal sample prepared with  $42.5$ wt \% cesium--pentadecafluoroctanoate (Cs--PFO) and  $57.5$ wt \% deuterated water (D$_{2}$O) \cite{boden1993}. Cesium nuclei ($^{133}$Cs) are quadrupolar spin systems with $I=7/2$ such that the dimension of the Hilbert space is $d=2I+1=8$. The experiment was carried out in a Varian 500 MHz - Premium Shielded (11.7 T) spectrometer at room temperature (25 $^{\circ}$C). A liquid NMR 5mm probe was used in this experiment. The Larmor frequency and quadrupolar coupling are 65.598 MHz and 7.7 kHz, respectively. The $\pi $-pulse time was calibrated as 25 $\mu $s. The spin--lattice  and spin--spin relaxation time, for  cesium nuclei, is $T_{1} \approx \ 320$ ms  and  $T_{2} \approx \ 4$ ms, respectively. The recycle delay time is $d_{1}$ = 1.8 s.

To describe a quantum state in NMR implementations, we use the density operator formalism representing the thermal equilibrium state, whose populations satisfy the Boltzman-Gibbs distribution. Theoretically, the density operator is represented by $\rho =\frac{1}{\mathcal{Z}}\exp\left[-\beta \mathcal{H}_{0} \right]$, where $\mathcal{H}_{0} = -\hbar   \omega _{L}  \mathbf{I} _{z}$ is the secular contribution of the NMR Hamiltonian and $\beta=\left(k_{B} T\right)^{-1}$, $k_{B}$ being the Boltzmann constant $T$ the room temperature. If the polarization strength is $\epsilon =\beta \omega_{L}\hbar/ \mathcal{Z} $, where $\mathcal{Z}$ is the partition function, and this factor has a value $\sim 10^{-5}$ then the density operator could be expanded to a first order approximation as  $\rho =\frac{1}{\mathcal{Z}}\mathbf{1}-\epsilon \rho_{0}  $, in which $  \rho_{0} = \mathbf{I}_{z}$ is called the deviation density matrix.

To initialize the quantum state, we transform $\rho_{0}$ to prepare a pseudo-nuclear spin coherent state (pseudo--NSCS) using the protocol of Ref. \cite{auccaise2012,araujo-ferreira2012}. The pseudo--NSCS is denoted as $\left| \zeta \left( \theta,\varphi \right) \right\rangle$, so the density operator is  $\rho =\left(\frac{1}{8}-\epsilon \right) \mathbf{1} + \epsilon \Delta \rho $, such that $\Delta \rho  \equiv \left| \zeta \left( \theta,\varphi \right) \right\rangle \left\langle \zeta \left(  \theta,\varphi \right) \right| $, for any  $0 \leq \theta \leq \pi$ and $0 \leq \varphi \leq 2 \pi$.

For our purpose, the angular parameters (polar $\theta$ and azimuthal $\varphi$) of a pseudo-NSCS were chosen in such a way that the phenomenon of bifurcation appeared as sharply as possible and therefore these parameters were fixed at a pair of initial conditions $\left| \zeta _{+}\left( \pi /4 ,\pi \right) \right\rangle $ and $\left| \zeta _{-}\left( 3\pi /4 ,\pi \right) \right\rangle $, where the $\zeta$--positive  ($\zeta$--negative) represents an initial condition at the north (south) hemisphere of a spherical phase space.

To sketch the classical bifurcation we use a control parameter $\Lambda$, such that $\omega_{Q}$ is maintained at a constant strength and  $\omega_{1}$ is varied from highest to lowest values. The different strengths of $\omega_{1}$ are quantified by the calibration of $\pi$--pulses at various elapsed times $t_{\pi}=$25, 30, 40, 50, 60, 100 $\mu$s or for the parameter $\Lambda=0.67, \ 0.81, \ 1.08, \ 1.35, \ 1.62, \ 2.70$. Once we have chosen the strength of $\omega_{1}$, the implemented pseudo--NSCS $\left| \zeta_{\pm} \left( \theta,\varphi \right) \right\rangle  $ is transformed by a $RF$ pulse that depends on the Hamiltonian in Eq. (\ref{squeezingRMNQ}) at various times $\tau_{k} = k \Delta \tau $ and $k=0,\ldots , 44$  with $\Delta \tau = 5 \mu$s. At each step, the pseudo--NSCS is tomographed by the quantum state tomography (QST) procedure \cite{araujo-ferreira2012,teles2007}. Next, the average value of the $z$--component of the spin angular momentum,  $\left\langle \textbf{I}_{z}\left(\tau_{k} \right)\right\rangle=\texttt{Tr}\left\{\Delta \rho\left(\tau_{k}\right) \ \textbf{I}_{z} \right\}$, is computed.

\begin{figure}[tbp]
\includegraphics[width=0.48
\textwidth]{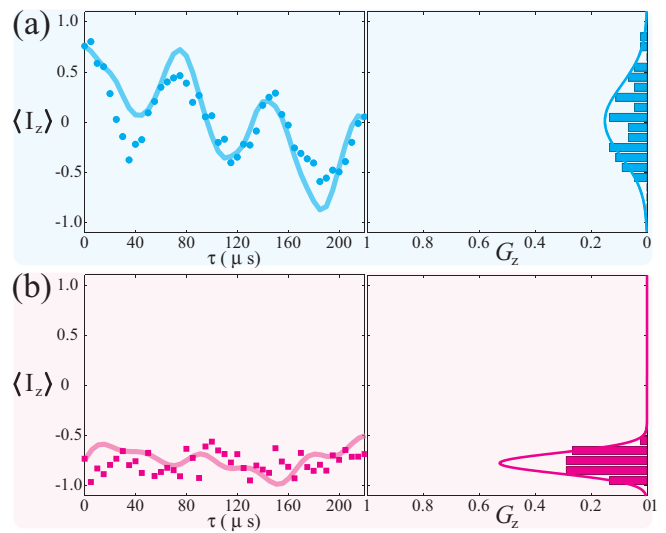}
\caption{(Color online) Experimental results (symbols) and numerical results (continuous lines) for the dynamics of mean average values of  $z$--magnetization (left) and a $z$--magnetization distribution (right).  Each value was calculated from the tomographed deviation density matrix at 45 different values of $\tau$. Results for (a) initial condition on north hemisphere and a linear regime, (b) initial condition on south hemisphere and a nonlinear regime.}
\label{fig:ValorMedioZ}
\end{figure}

\begin{figure}[tbp]
\includegraphics[width=0.48		
\textwidth]{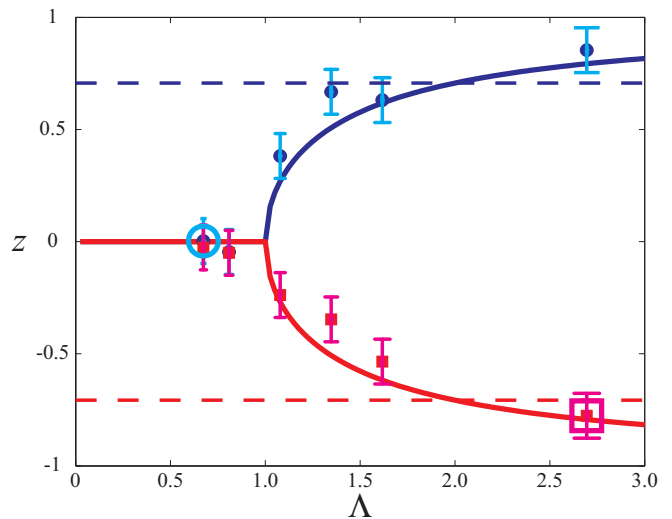}
\caption{(Color online) Experimental results (symbols) of bifurcation phenomena in spin systems which are described by the JJ--model. Two initial conditions are denoted by dashed lines, which correspond to $\left|\zeta_{+}\right\rangle$  and $\left|\zeta_{-}\right\rangle$ pseudo-NSCS. The spin system can be driven under different regimes:  a linear $\left(0<\Lambda<1\right)$ and a non--linear  $\left(1<\Lambda\right)$. Theoretical predictions (solid lines) are sketched by using $\pm \sqrt{1-1/\Lambda}$,  whose positive (negative) sign corresponds to the north (south) hemisphere of a spherical phase space. The cyan circle (magenta square) indicates that the data was calculated from experimental results explained in  Fig. \ref{fig:ValorMedioZ}(a) (Fig. \ref{fig:ValorMedioZ}(b)). The error bars are bounded at 10 \% of the maximum value of the time averaged $z$--magnetization, which is $\left|\pm 1\right|$.}
\label{fig:BifurcationCurves}
\end{figure}

In Fig.  \ref{fig:ValorMedioZ} we show the experimental results for $z$--magnetization and its distribution for initial conditions on the north (south) hemisphere  of the spherical phase space under the linear (nonlinear) regime.
On the left of Fig. \ref{fig:ValorMedioZ}(a) are shown experimental (dots) and numerical (solid line) results \footnote{We refered as numerical results when they are generated transforming the tomographed initial pseudo-NSCS by an operator that depends on Eq. (\ref{squeezingRMNQ}).} for  $\left\langle \textbf{I}_{z}\left(\tau\right)\right\rangle$, for the initial condition on the north hemisphere under the linear regime, but not too far from the nonlinear regime, with approximately $\Lambda\approx 0.67$.  The beats and fast decay is a typical signature of an intermediate regime. On the right of Fig.  \ref{fig:ValorMedioZ}(a) we display a histogram of the experimental values of $\left\langle \textbf{I}_{z}\left(\tau\right)\right\rangle$ and the Gaussian distribution $G_{z}\left(\left\langle     \textbf{I}_{z}            \right\rangle\right)=G_{0} \exp\left[-\left( \left\langle \textbf{I}_{z}\right\rangle-z  \right)^{2}/\sigma^{2}_{z}\right]$ has been calculated and drawn. The parameter $z$, referred to as the temporal mean $z$--magnetization, corresponds to the mean value of $\left\langle \textbf{I}_{z}\right\rangle$ over 45 different elapsed time points. Similarly, the parameter $\sigma_{z}$ is the well--known standard deviation. The main information extracted from  $G_{z}$ is the $z$--value, which depends on $\Lambda$, and this parameter indicates the stage of bifurcation.

On the left of Fig. \ref{fig:ValorMedioZ}(b) we show experimental (squares) and numerical (solid line) results for  $\left\langle \textbf{I}_{z}\left(\tau\right)\right\rangle$ in the nonlinear regime, satisfying  $\Lambda = 2.7 > 1 $.  The smooth beats are almost completely attenuated and the decay is slower than in the linear regime. This happens because $\omega_{1}$ is weaker than $\omega_{Q}$.  On the right of Fig.  \ref{fig:ValorMedioZ}(b), there is a histogram for experimental values of $\left\langle \textbf{I}_{z}\left(\tau\right)\right\rangle$ and a Gaussian distribution $G_{z}\left(\left\langle     \textbf{I}_{z}            \right\rangle\right)$.

In Fig. \ref{fig:BifurcationCurves} we present the experimental results (symbols) and theoretical prediction (solid lines) of the temporal mean $z$--magnetization for initial conditions on the north hemisphere (dots) and south hemisphere (squares). The cyan circle (magenta square) indicates that the data was calculated from experimental results explained in  Fig. \ref{fig:ValorMedioZ}(a) (Fig. \ref{fig:ValorMedioZ}(b)). We observe that the experimental results match the theoretical prediction of bifurcation. To explain this phenomenon in a nuclear spin system, we need to remember that the eigenstates of the secular Hamiltonian are $\left|m\right\rangle$, with $m=-I,-I+1,\ldots,I-1,I$, and the operator that depends on Hamiltonian (\ref{squeezingRMNQ}) transforms each eigenstate. In the linear regime, the $RF$ term imposes the dynamics of spins, maintaining the spin system under a superposition of the $\left|m\right\rangle$ eigenstates, leading the quantum state from $\left|\zeta_{+}\right\rangle$ to $\left|\zeta_{-}\right\rangle$ and vice-versa. This is analogous to what happens in the tunneling phenomenon of a symmetrical trap of a two--mode BEC \cite{albiez2005,gati2007,raghavan1999,sorensen2001May,hines2005,kellman2002,zibold2010,shchesnovich2009,diaz2010Feb,diaz2010Jun,rubeni2012} or what happens in the description of the JJ--model in superconductivity, called the plasma oscillation regime \cite{gati2007}. On the other hand, in the nonlinear regime, the nonlinear term decides the behaviour of the system. In this case, from the basic principles of Quantum Mechanics, we know that $\mathbf{I}_{z}^{2}\left|\pm m\right\rangle = \hbar m^{2}\left|\pm m\right\rangle$; for $\left|+m\right\rangle$ and $\left|-m\right\rangle$ there is a degeneracy which drives the spin system to the bifurcation phenomenon, because the initial condition for the north (south) hemisphere is retained by the operator that depends on Hamiltonian (\ref{squeezingRMNQ}) and the spin system is induced to precess aligned parallel (antiparallel) to an effective magnetic field which is aligned along the $z$--axis of a reference frame. Similarly, in a two--mode BEC, this regime corresponds to the self--trapping regime \cite{gati2007}.

Finally, we draw our conclusions: we performed a classical bifurcation in a nuclear spin system that is described and interpreted by the JJ--model. The spin scenario coincides with other experimental schemes commonly named as the symmetric double--well trap of two--mode BEC, with the possibility of extending to the asymmetric case. We take advantage of the main physical property of  a lyotropic liquid crystal sample, which is the collective order in the presence of a magnetic field. This inspired us to explore the nonlinear regime to study squeezing processes, which are currently being developed in our laboratory.

\begin{acknowledgements} 
The authors acknowledge financial support from CNPq, CAPES, FAPESP, and FAPERJ.  This work was performed as part of the Brazilian National Institute of Science and Technology for Quantum Information (INCT-IQ). We also acknowledge P. Judeinstein for Cs--PFO samples and thank Angela Foerster for meaningful discussions and suggestions. 
\end{acknowledgements} 
%---------------------------------------------------------------------------------------------------------

\end{document}